# Authenticity as Aesthetics: Enabling the Client to Dominate Decision-making in Co-design

Jingrui An [1*]

1. Eindhoven University of Technology, Eindhoven, The Netherlands

## Abstract

This paper revises aesthetics theory through the lens of authenticity and investigates practical applications using a co-design approach. We encourage designers to include ordinary clients as co-creators in the co-design process, guiding them in expressing their aesthetics, values, and preferences while stimulating their creativity. This paper proposes a bespoke design process framework for authenticity aesthetics that incorporates empathy, defining, ideating, prototyping, and testing. This framework delineates the roles and responsibilities of clients and designers at different phases and highlights evolving material mediums that enable their communication. The paper concludes by reflecting on consumerist aesthetics, advocating for designers to focus on the insights of ordinary clients, design for their authentic uniqueness, and recognize the broad prospects of bespoke design methods.

## 1. Introduction

In the experience economy [1], aesthetics permeates data, artifacts, and service use, influencing individuals' experiences and choices. This emphasis on aesthetics has led to a focus in design discourse on integrating connectivity into design and empowering individuals to express their values and aesthetics [2]. However, the rise of consumerist aesthetics, which encourages a relentless pursuit of novelty, can lead to a loss of personal aesthetics and reinforce consumerism [3]. In the tourism industry, this trend manifests as locals copying successful models or creating artificial attractions for short-term gains, ultimately sacrificing authenticity and threatening local socio-cultural sustainability [4].

To address the loss of personal aesthetics and cultural authenticity, we advocate for empowering individuals who aspire to create tourist experiences but lack design skills. By collaborating on personalized designs, these individuals can infuse their unique values and preferences into the experiences they create. This approach aligns with the growing recognition of the importance of co-creation in design [5]. However, despite its user-centered approach, designers' aesthetic preferences and popular notions still influence the decision-making process, potentially overlooking individuals' aesthetic values [6] and even promoting stereotypes or biases [7].

Our research acknowledges that individuals are the experts in their own lives [8]. Therefore, designers should actively encourage clients (i.e., end users or service providers) to express their unique values and aesthetic preferences, even if they lack professional design knowledge. By making authenticity the central aesthetic principle, we shift the focus away from superficial trends and towards authentic experiences and narratives that shape individual perspectives. This study proposes a co-design framework based on authenticity to facilitate this shift.

## 2. Background and Related Works

### 2.1. Authenticity as Aesthetics

As an intrinsic value, aesthetics is the perception that certain things hold inherent worth, recognized by appreciating something for its own sake and acknowledging its uniqueness [9]. Existentialist aesthetics further posits that humans are the means through which things manifest, and art primarily aims to express uniquely human qualities intentionally [10]. Hekkert [11] identifies four principles of aesthetic pleasure: 1) *maximum effect for minimum means*, enabling efficient system operation and quick decision-making; 2) *unity in variety*, reducing attention resource allocation through relationship extraction; 3) *most advanced, yet acceptable*: people prefer new, unfamiliar, and original things while retaining typical features for adaptation; 4) *optimal match*: multimodal stimuli provide consistent information across senses. The aesthetic significance of artifacts, then, derives from their ability to facilitate meaningful engagement in essential activities [12]. Authenticity, a core element of everyday aesthetics, is evident in an object's uniqueness, irreplaceability, and irreducibility, allowing it to maintain value despite fleeting trends [3].

Existential-phenomenological authenticity has two main components [13]. First, authentic behavior or communication is seen as the individual's spontaneous creation—willingly enacted, owned, and self-recognized. Second, someone acts under their core beliefs. Su and Stolterman [14] proposed two key concepts for conveying authenticity: *spheres of existence* and *indirect communication*. The *spheres of existence* encompass self-awareness, critical reflection on personal goals and values, and responsibility for actions, akin to being authentic to oneself. *Indirect communication* involves expressing one's existence through cues others can interpret, revealing both positive and negative aspects of existence. However, authenticity in tourism can be artificially constructed, projecting the service provider's image onto tourists, known as "staged authenticity" [15]. While Massi et al. [16] emphasized prioritizing a smooth customer journey for authentic experiences, we believe design activities can intentionally highlight individuals' personalities and preferences, fostering a harmonious match between individuals and their natural aesthetic environment. This approach moves beyond staged authenticity towards a more genuine expression of individual identity and values.

Based on these theoretical foundations, this paper proposes a design process that prioritizes authenticity, enabling clients (end users or service providers) to express their aesthetics. Realizing authenticity's value and its relationship to personal experiences empowers ordinary clients. We summarize the aesthetic principles that delight clients from an authentic perspective: 1) integrating design with other existing related experiences in the environment to minimize disruption to the users' experience journey; 2) embodying the client's unified self-awareness, preferences, and values; 3) presenting novel experiences in a manner consistent with the client's recognized habits; and 4) ensuring that stimuli received by different senses are harmonious and matched.

After discussing the principles of aesthetics that satisfy authenticity, we will look at how design can help ordinary clients express authenticity.

## 2.2. Client-driven Co-design Decisions

Product and service design in the experience economy focuses on personalized user experiences rather than functionality. Co-design, which prioritizes user participation, is important for achieving this goal [17]. Cross [18] contended that genuine participation in design entails "Do-it-yourself," enabling participants to create their own rules and engage those without design skills. Lee [6] proposed a three-phase co-design process: preference, planning, and processing. This process prioritizes user involvement and creativity throughout the design journey. Participants express their preferences using customized tools during the preference phase. In the planning phase, they collectively define their design brief and engage in self-study exercises. Finally, the processing phase transforms abstract design tools into interactive games, providing participants with a hands-on design experience.

However, researchers acknowledge that co-design often produces abstract results [19], making them challenging to apply in real-world situations. Bespoke design tailors products and services to specific users or groups and produces real-world solutions or research products valued by researchers. Some researchers have detailed bespoke design processes for specific cases. For instance, Gaver et al. [20] bespoke the Prayer Companion for nuns, which involves *contextual inquiry, initial design development, design refinement, and living with the Prayer Companion*. Boucher and Gaver [21] described a process for designing Datacatchers for travelers that involves *background and initial concept, conceptual direction (e.g., form and affordances, interaction and participation with data), and refinement and production (product and form sketches)*. Wallace et al. [22] described a process for bespoke the interactive Self-Reflector for a shop, generally including *conceptual direction, design ideation, form and aesthetics, and interaction*. Desjardins et al. [19] proposed the "bespoke booklet" method for critical design ideation, enabling users to express their needs and preferences through sketches and text.

Odom et al. [23] referred to these research-driven artifacts as "research products" requiring four qualities- *inquiry-driven*, *finish*, *fit*, and *independent*. Research inquiry [23] guides most researchers' bespoke design efforts, focusing on situated knowledge through practice [24]. However, despite similarities in the bespoke process, we couldn't find general frameworks to help designers with bespoke design. Most bespoke design studies do not disclose design decision criteria. We believe that designers' judgments about context and interactional appropriateness may overlook clients' potential value from an authenticity perspective.

Furthermore, we also acknowledge that designers' aesthetic preferences or biases can influence initial design concepts. Designers provide materials, such as brainstorming stimuli [25], mood boards [26], and technology probes [27], based on their understanding of client needs [28] to stimulate clients' further expression. However, individuals are experts in their own lives, and designers may find it difficult to empathize fully and guide clients unbiasedly in the early stages.

Recent research has explored the potential of AI-generated content (AIGC) to inspire design ideas [29]. Huang et al. [30] proposed a framework for AIGC-enabled design processes, including empathy (user studies), define (feature description), ideate (text-to-AIGC image and concept sketches), prototype, and test. AIGC's content generation from existing works improves concept visualization efficiency [31], potentially reducing designers' aesthetic biases. However, AIGC is not exempt from its own biases [32], necessitating a balanced approach that leverages its broad form and aesthetic sample advantages while relying on designers' empathy to address potential biases.

## 3. Bespoke Design Process for Authenticity Aesthetics

Despite the numerous advantages of bespoke design, involving non-designers in technically demanding processes may undervalue professional designers' expertise [33] and overwhelm users [5]. To address this, we propose a "Bespoke Design Process for Authenticity Aesthetics" that designers can apply when prioritizing authenticity. This process integrates principles for designing for authenticity [14, 13], existing bespoke design processes (e.g., [20, 21, 22, 19]), and leverages AIGC to enhance efficiency, reduce communication costs, and mitigate biases. We build upon Huang et al. [30]'s design process—*empathy*, *define*, *ideate*, *prototype*, and *test*—which encompasses previous bespoke processes and design thinking considerations (Figure 1).

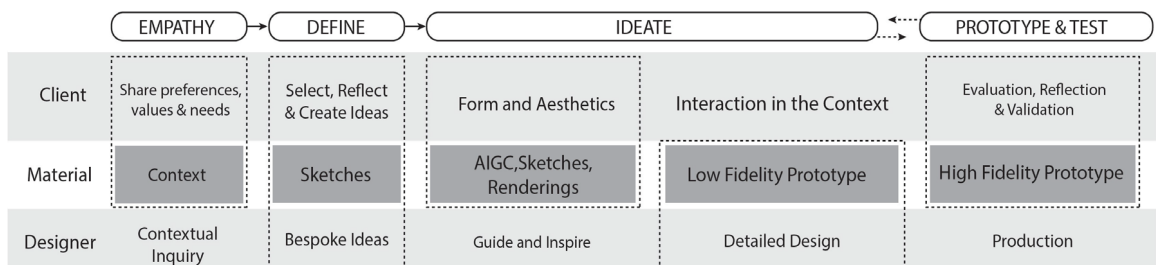

**Figure 1:** Bespoke Design Process for Authenticity Aesthetics, the dashed square box in the diagram represents the dominant role in the different steps.

## 3.1. Empathy

The empathy phase enables designers to understand clients' authentic preferences, values and needs [13]. Designers immerse themselves in their clients' lives, using observational and interview methods to identify the aesthetic unity within the context [11], often documenting through photography. Simultaneously, clients share personal anecdotes revealing their preferences and needs [34] and assist in capturing details that designers might miss [19]. These on-site photos and narratives facilitate a comprehensive understanding of the design context and client needs. By guiding clients through self-disclosure [3], designers establish initial empathy during this phase.

## 3.2. Define

In the define phase, design concepts are articulated and integrated into the experience journey, enabling end users to perceive aesthetic pleasure more quickly and easily [11]. Designers communicate with clients through bespoke design concept sketches (e.g., [19]). Design sketches describe concepts' functions, interactions, and intended experiences in real-world settings. Initial sketches help clients envision future possibilities [19]. Clients choose their preferred experiences [35] and actively shape the final concepts, ensuring their voluntary formulation and self-approval [13].

Each concept aims to customize the end user's experience through interaction design touchpoints [36]. Meanwhile, positioning interactive touchpoints throughout the consumer journey is important [37]. To maintain sensory consistency, designers should integrate new touchpoints with adjacent ones in the experience journey [11]. Designers can now assess design concepts' expressive intent, general aesthetic preferences, and interaction methods.

## 3.3. Ideate

The ideate phase consists of two steps: *Form and aesthetics*, and *Interaction in the context*.

### 3.3.1. Form and aesthetics

In our proposed framework, the client takes the lead in expressing design concepts of form and appearance [20]. Leveraging the power of AI-generated content (AIGC) as material, clients can either provide prompts or collaborate with large language model platforms like ChatGPT to generate detailed concept renderings through AIGC platforms such as Midjourney or Stable Diffusion. This AIGC-driven approach offers a significant advantage over traditional design sketches, enabling clients to intuitively perceive aesthetic forms more efficiently and completely [30]. Based on the AIGC results, clients can refine their requirements, iteratively working towards outcomes that meet or exceed their expectations. The designer guides and inspires the client to refine concept descriptions by using their professional expertise to ask critical questions and facilitate flexible AIGC platform use. AIGC's outputs may be biased or inaccurate, and the structure or materials of outcomes may not be practical. The designer should use their expertise to enhance concepts [38, 22]. The designer turns AIGC outcomes into practical sketches and renderings, ensuring fit within the intended setting [23]. These materials simplify communication and enable detailed client feedback.

### 3.3.2. Interaction in the context
To balance novelty with end-user habits, low-fidelity models like tangible foaming models or Figma interactive demos simulate interaction forms and potential issues [11]. Service providers who want attractive and engaging designs will find this novelty relevant. Interactive prototypes let clients experience size and texture that renderings cannot. This embodiment lets them provide feedback on requirements that are not apparent from visuals.

During this stage, the designer completes the design's appearance, texture, and interaction methods. Furthermore, designers have to collect specific information about the interaction interface from the client and authorization data relevant to the interaction [22]. This ensures the design is consistent with the client's tastes and practical needs.

By combining these steps, the ideate phase ensures that design concepts are aesthetically pleasing and practical, aligning with the clients' authentic preferences and needs.

### 3.4. Prototype &Test

The Prototype&Test phase aims to create high-fidelity, independently operational products [23]. Designers develop final products based on client feedback and prototype performance [21]. Next, the bespoke design should be tested against commercial standards in real-world scenarios. The products should function independently in real-world settings without client or designer assistance [23], as well as expressing clients' preferences, personalities, and value propositions. For academic designers, bespoke design artifacts that integrate with the client's authentic usage context should also be inquiry-driven with specific research questions.

In the next section, we will demonstrate the *Bespoke Design Process for Authenticity Aesthetics* through a case study of custom design for Airbnb hosts. This will show how we collaborated closely with hosts to create artifacts tightly connected to Airbnb aesthetics.

## 4. Discussion

This research examines co-design through the lens of "authenticity as aesthetics." We integrated theories of aesthetics and authenticity to critique the "design savior" complex [39], wherein designers assume they have superior skills to assess and solve unfamiliar problems quickly. Drawing upon existing bespoke design processes, we propose the *Bespoke Design Process for Authenticity Aesthetics*, offering insights into empowering individuals to express their unique values and aesthetics, thus counteracting the homogenizing effects of consumerism. This study explores the shift from a market-driven to a human-centered era in design and the broader phenomenon of individuals without formal design training engaging in design [17].

Our process creatively combines existing bespoke design methods (e.g., [19, 20, 21, 22]), summarizing the entire workflow from initial inquiry to final product delivery. Previously, the bespoke design was frequently viewed as a process for developing technology probes [27] or material speculations [40], primarily motivated by research questions. Consequently, these studies lacked a standardized process and mostly focused on describing the research context rather than considering design decisions. We've combined these processes into an overall design process that clearly defines each step's theme, purpose and materials. This structured approach not only streamlines the design process but also helps to control material and time costs.

In the era of experience economy and rapid advancement of artificial intelligence, clients are transitioning from stakeholders to co-creators, transferring their non-designer experiences to designers. Within this framework, designers are not excluded from the process; instead, they empower individuals by providing tools for ideation and expression (e.g., Bespoke booklet [19], AIGC) [41], facilitating clients' creativity at every step [17], and drawing upon past experiences to alert clients to potential issues they might overlook.